\documentstyle[prl,aps,epsfig,floats,twocolumn]{revtex}

\def\ni{\noindent}
\def\br{\vec{r}}
\def\bg{\vec{g}}
\def\bs{\vec{s}}
\def\bt{\vec{t}}

\def\bR{\vec{R}}

\def\bff{\vec{f}}

\def\brho{\vec{\rho}}
\def\bnabla{\vec{\nabla}}

\def\hC{{\hat C}}

\def\hP{{\hat P}}

\def\hQ{{\hat Q}}

\def\hS{{\hat S}}
\def\sig{\sigma}  
\def\hsig{{\hat \sigma}}
\def\heps{{\hat \epsilon}}

\def\cA{{\cal A}}

\def\bcF{\vec{{\Phi}}}

\begin{document}
\draft


\title{Stress transmission in planar disordered solid foams}
\author{Raphael Blumenfeld}

\address{Polymers and Colloids, Cavendish Laboratory, Madingley Road,  
Cambridge CB3 0HE, UK} 
\maketitle 
\date{\today} 
\maketitle 

\begin{abstract} 
Stress transmission in planar open-cell cellular solids is analysed using a recent theory developed for marginally rigid granular assemblies. This is made possible by constructing a one-to-one mapping between the two systems. General trivalent networks are mapped onto assemblies of rough grains, while networks where Plateau rules are observed, are mapped onto assemblies of smooth grains. The constitutive part of the stress transmission equations couples the stress directly to the local rotational disorder of the cellular structure via a new fabric tensor.   
An intriguing consequence of the analysis is that the stress field can be determined in terms of the microstructure alone independent of stress-strain information. This redefines the problem of structure-property relationship in these materials and poses questions on the relations between this formalism and elasticity theory. The deviation of the stress transmission equations from those of conventional solids has been interpreted in the context of granular assemblies as a new state of solid matter and the relevance of this interpretation to the state of matter of cellular solids is discussed. 
 
\end{abstract}
\pacs{62.20.-x, 61.43.Gt, 81.40.Jj}
\narrowtext

{\bf I. Introduction}
\smallskip

Planar foams consist of partitions of space by irregular cells and are observed in many natural and man made systems. Their ubiquity in nature is often due to evolutional processes that have led to an optimization of mechanical or transport properties against weight constraints. Much effort has been invested in attempts to follow the wisdom of Mother Nature and develop materials with biologically inspired structures. This effort has resulted in a growing body of knowledge on the characterization of the structure as well as on understanding of the geometry of the cells in terms of the nonequilibrium dynamical processes that generate them. Essential to this entire field is a basic understanding of the relations between the internal structural characteristics of such media and their macroscopic properties \cite{vincent}. Such an understanding would be of use to a wide range of applications, such as design of high-perfrmance and smart materials, bio-artificial technology, and the food insustry, to name only a few. A corner stone in the study of cellular materials is the work of Gibson and Ashby \cite{ashby} who employed various approximations on ordered cellular structures to estimate the mechanical properties of cellular networks in general. While that work has significantly improved the theoretical basis for stress calculations, it has failed to provide a rigorous relation between the local microstructure in general disordered systems and the local stress field. It has become increasingly evident in recent years that the lack of such a first principle relation stands in the way of progress both on the theoretical and the technological fronts. 
 
It is conventional to relate constitutive structural data in solids to macroscopic properties by integration over microscopic variables \cite{HS}\cite{Beran}\cite{Christensen}\cite{Willis}\cite{Torquato}. In this approach one starts from a given constitutive stress-strain relation on a judiciously chosen microscopic lengthscale and coarse-grains this relation until it applies to a suitable macroscopic lengthscale. This paradigmatic approach has been automatically adopted in recent era for cellular solids but the main problem has always been the translation of the local cellular structure, which can be arbitrary in general systems, into local elastic properties. The question whether it is really necessary to go down this route at all has not been seriously addressed. The results in this paper suggest an alternative approach that obviates the problem of finding the elastic constants altogether. 
 
On the conceptual level, the problem is the following: Suppose a cellular system, whose structure is known in every detail, is in mechanical equilibrium under a set of forces $\bg$ acting on its boundaries. Suppose also that the properties of the material constituting the network are well characterized. The questions are: (i) how to describe the discrete field of forces transmitted by the cell walls as a continuous stress field? (ii) what are the equations that govern this field (the stress transmission equations)?  and (iii) how do different microstructures affect the local stress?  
In $d$ dimensions it takes $d^2$ conditions to determine the $d\times d$ stress tensor $\sigma$. Force balance gives $d$ conditions,   
 
\begin{equation}
\bnabla\hsig = \bg
\label{eq:Ai}
\end{equation}
and blanace of torque provides further $d(d-1)/2$ conditions - one for each axis of rotation,  
 
\begin{equation}
\sigma_{ij} = \sigma_{ji}
\label{eq:Aii}
\end{equation}
Equations (\ref{eq:Ai}) and (\ref{eq:Aii}) give altogether $d(d+1)/2$ relations and therefore additional $d(d-1)/2$ equations (one in $d=2$ and three in $d=3$) are required to solve for the stress components. These 'missing' equations must consist of information about the constitution of the material - the constitutive equations. In conventional elasticity theory, this second set of equations is obtained by imposing compatibility conditions on the strain field and relating the stress to the strain through constitutive relations that are independent of the balance conditions. While equations (\ref{eq:Ai}) and (\ref{eq:Aii}) couple the stress to the external loading, the constitutive equations relate it to the properties of the medium and determine how the stress fluctuates due to spatial variations of these properties. In cellular systems, these variations are governed by the microstructure and so it seems that there is no escape from the problem of structure-property relations. Yet, what is the property that the structure needs relating to? The basic issue is not how to relate the structure to elastic constants, which we can then plug into the comaptibility relations. Rather, it would suffice to find $d(d-1)/2$ relations that couple the structural information directly to the stress field.  
 
A development in the study of granular systems has provided an unexpected key for progress on this problem. In a recent paper, Ball and Blumenfeld (BB) \cite{BB} have formulated a new theory for stress transmission in two dimensional marginally rigid granular assemblies \cite{soft}. In particular, they have derived a constitutive equation that couples the local stress to the local geometry. Marginal rigidity is a new state of solid matter, whose existence has been verified experimentally \cite{BEB}. In this state the number of intergranular contacts in the system allows to determine the microscopic force field from statics alone and therefore the macroscopic stress is independent of the stress-strain data. BB have further shown that the new theory predicts that the symmetric divergence-free stress field is the Airy stress function \cite{muskh}, recovering a result of conventional elasticity theory and so confirming the redundancy of the stress-strain relations in marginally rigid systems. 

The aim of this paper is to  to adapt the BB formalism to planar trivalent cellular structures and derive a constitutive relation between the microstructure of a model of such systems and the stress field that develops in them under external loading. To this end, an exact mapping is formulated between cellular networks and granular assemblies at the marginal rigidity state.  
By decoupling the stress from the strain, the new theory obviates the problem of structure-property relationship in the traditional sense and provides a direct structure-field relationship. In the context of granular systems, marginal rigidity is a new state of solid matter and this begs the question whether cellular solids and solid foams are also in this state.  
 
The paper is structured as follows: Section II presents an argument that leads to the conclusion that the stress field in cellular networks at equilibrium can be determined from the loading and the microstructure alone, independent of any strain-stress information. Section III formulates a one-to-one mapping between planar trivalent cellular networks and marginally rigid assemblies of grains. Section IV describes the derivation of the constitutive equation and section V concludes with a brief discussion of the results and some of their implications.    

\bigskip 
\ni {\bf II. The stress field can be determined independently of stress-strain relations} 
\smallskip 
 
In this section it is shown from first principles that the stress field can be determined from statics alone. The argument is slightly different for general trivalent networks, where the cell walls meet at arbitrary angles, and for Plateau networks. Therefore these two cases will be treated separately throughout the paper. 
In the following we model a cellular solid by a collection of vertices at given positions in the plane, which are connected by cell walls whose thickness is much smaller than the size of a cell (a dry foam). These cell walls may be straight lines or curved as a consequence of the process that has given rise to the microstructure and with which we are not concerned in the present discussion.

\smallskip  
{\bf (a) General networks} 
\smallskip 
 
Consider a planar open-cell trivalent cellular network that is at mechanical equilibrium under a set of external forces, $\bg$, acting on its boundaries and which consists of $N$ cells. The loading gives rise to tension (or compression, which will be regarded in the following simply as negative tension) in the cell walls and to torques on the vertices of the network. The former are balanced by forces along the cell walls, while the latter must be balanced by forces that are normal to the walls. By Euler's theorem \cite{sixedge} for $N\gg 1$ a cell has on average six neighbors and, since each cell wall is shared by two cells, there are altogether $3N$ cell walls. Each vertex in the network is the meeting point of three walls and each wall connects two vertices and so there are exactly $2N$ vertices. This enumeration neglects boundary effects, which introduce corrections of order $O(1/N)$, but which can be readily included into the argument that follows and therefore do not affect its generality. 
 
\begin{figure} 
\centerline{\psfig{file=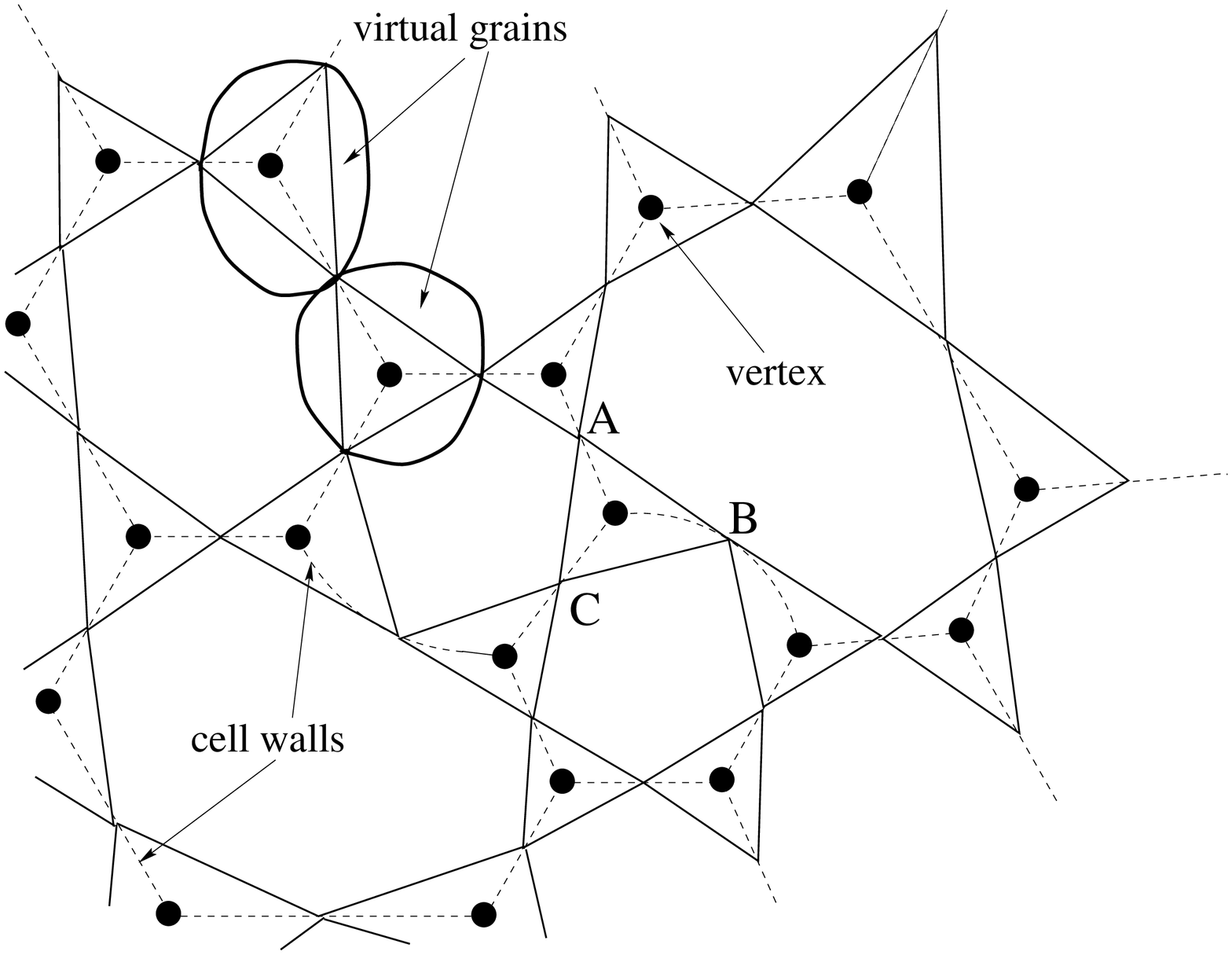,height=6.5cm}} 
\caption{ 
The mapping from a trigonal cellular network onto an assembly of grains under 
external loading. The original cellular network is shown in dashed lines. 
The midpoints of the cell walls are joined to form triangles around each vertex (black circle). 
The grains of the equivalent granular assembly are centered on the vertices with the corners of 
the triangles representing contacts between neighboring grains. Two such virtual grains in contact 
are shown.
} 
\end{figure} 

Let us draw around each vertex a triangle whose corners are located at the midpoints of the walls that branch from it (e.g. triangle ABC in figure 1) and let us imagine that the edges of these triangles are made of rigid struts. Each  of the 2N triangles is in contact with three neighbors and the contact points are considered rigid so that the triangles cannot rotate relative to each other around them. Two triangles in contact exert a force on each other whose two components are defined to be the tension along the wall and the force normal to the wall that contributes to balancing the torques on the vertices due to the external loading. Within this description there are no torques on the contact points (which may result, for example, from a finite thickness of the cell walls). Thus defined, the triangles form a rigid frame that transmits the same forces and moments as the original cellular structure. There are $3N$ contact forces and therefore, to determine the intercellular forces, one needs to solve for $6N$ unknowns altogether.  

How many equations are available to determine these? The triangles are at mechanical equilibrium and so each must obey two equations of force balance and one equation of torque balance. Multiplied by the $2N$ vertices, this gives $6N$ equations, exactly enough to determine all the forces. The balance equations depend only on the details of the microstructure and on the external loading. It follows that the intercellular forces can be uniquely determined from statics alone.  
 
\smallskip 
{\bf (b) Plateau networks} 
\smallskip 
 
In Plateau trivalent networks the cell walls meet at each vertex at angles of $2\pi/3$, a structure that abounds in nature and man-made materials. In contrast to the general case, force balance at every vertex dictates here that all the cell walls are under the same tension. This leaves the normal force on each cell wall as the only unknown and the number of degrees of freedom is reduced from $6N$ to $3N$. We use the same construction as in figure 1, but instead of considering the frame of triangles, let us regard it as consisting of a set of connected rigid polygons, each enclosed inside a cell. These polygons connect to each other at the same points that the triangles do, namely, at the midpoints of the cell walls. If we define the forces transmitted between every two neighboring polygons as the forces normal to the cell walls at the midpoints, then this rigid frame supports the same stresses as the original system. Each of the $N$ polygons has to obey three balance equations, two of force and one of torque, which gives altogether $3N$ equations, exactly enouch to determine all the forces.  
 
So, the intercellular forces in both general and Plateau networks can be derived from statics alone and therefore these are isostatic systems. Now, the macroscopic stress is just a coarse-grained continuous description of the discrete field of forces on the cellular scale. It follows that since the latter can be determined without reference to compliance then so can the continuous stress field. It is evident then that the second set of equations that complements the balance conditions must arise from the structural characteristics alone.  
This observation has a significant ramification: regardless of the precise form that the consitutive equations take, the fact that the stress field equations involve no information on the stress-strain response means that the stress transmission in cellular solids is governed by {\it physical laws that are different than in conventional elastic solids} \cite{deform}. 
This basic realization appears to have been neglected in the large body of literature on solid foams. Most of the work in the field employs stress-strain relations, thus linking the stress to the displacement that occurs as the system approaches mechanical equilibrium. The fundamental issue underlying the present analysis is that in planar cellular solids the stress can be decoupled from any deformation. Thus, after all displacement has stopped, the stress can be calculated from the microstructure and the boundary conditions alone.  
 
\bigskip 
\ni {\bf III. Mapping cellular networks onto granular assemblies} 
\smallskip 
 
A planar trivalent foam can be characterized by the positions of all its vertices, $v_i$, a connectivity matrix that specifies which vertices are connected and the curvatures of the connecting walls. The cell walls tesselate the system into a set of cells, $c_i$. It is useful again to discuss separately the mappings of general and Plateau networks. 
 
\smallskip 
{\bf (a) General networks} 
\smallskip 
 
Let us consider again the construction shown in figure 1 and note that, although the cell walls may be curved, the edges of the constructed triangles are always straight. The above enumeration of equations versus degrees of freedom implies that one needs $2N$ rigid bodies and therefore suggests to center the equivalent grains on the vertices. The contacts between neighbor grains are at the midpoints on the cell walls between them where the respective triangles touch. The force $\bff_{vv'}$ that is transmitted at the contact between the grains on vertices $v$ and $v'$ is defined to be identical to the cell wall at this point. Namely, it consists of the tension along the wall, $T_{vv'}$, and the orthogonal component $F_{vv'}$ that assists in balancing the torques on vertices $v$ and $v'$ \cite{linetopoint}. The grains are considered to be static under the local forces applied on them (i.e., of sufficient rigidity to support the forces without deformation), and of sufficient roughness to prevent slip at any of the contacts under the intergranular forces. Note that the sides of triangle $v$ connect the contact points of the virtual grain centered on vertex $v$.  
 
\smallskip 
{\bf (b) Plateau networks} 
\smallskip 
 
Using again the construction of figure 1, let us now place a virtual grain inside every cell rather than around the vertices. The reason is that this system should provide $3N$ balance equations, which can only come from $N$ rigid bodies and $N$ is exactly the number of cells. The grains are again chosen to be sufficienly rigid to support the forces and the contacts between neighboring grains are at the midpoints of the cell walls. At the contact points the surfaces of the grains are presumed to be normal to the $3N$ unknown forces, and tangent to the wall directions. In contrast to the general case, the grains are supposed to be perfectly smooth and so incapable of supporting tangential forces at the contacts. Thus defined, the cell wall forces are mapped onto the normal forces that neighboring smooth grains exert on each other. The mean coordination number is six per grain, each of which is then associated with one unknown - the force normal to the surface.  
 
\smallskip 
 
Before we continue it is worthwhile to comment that there is a marked difference between real cohesionless granular assemblies and those described above. Real grains can only support compressive forces and tensile forces take them out of mechanical equilibrium. In contrast, cell walls can sustain either direction of forces. Correspondingly, the virtual assemblies described here which cellular networks map onto, can also have either type of forces between the virtual grains \cite{MaxwellA}. Although the original BB formalism has been developed for real systems under compressive forces, it can readily accommodate an overall change of sign into tensile forces.  

\bigskip 
\ni {\bf IV. The missing constitutive equation} 
\smallskip 
 
This section derives the constitutive equation by adapting the BB theory to the context of cellular solids. The derivation follows three main steps: First, the intercellular forces are described in terms of a new reduced set of forces that satisfies the force balance conditions. Second, the new forces are represented in terms of a continuous force field. The third and final step is to describe the stress as a function of the continuous field and apply the requirement of torque balance to obtain the missing constitutive equation.  
 
\begin{figure}
\centerline{\psfig{file=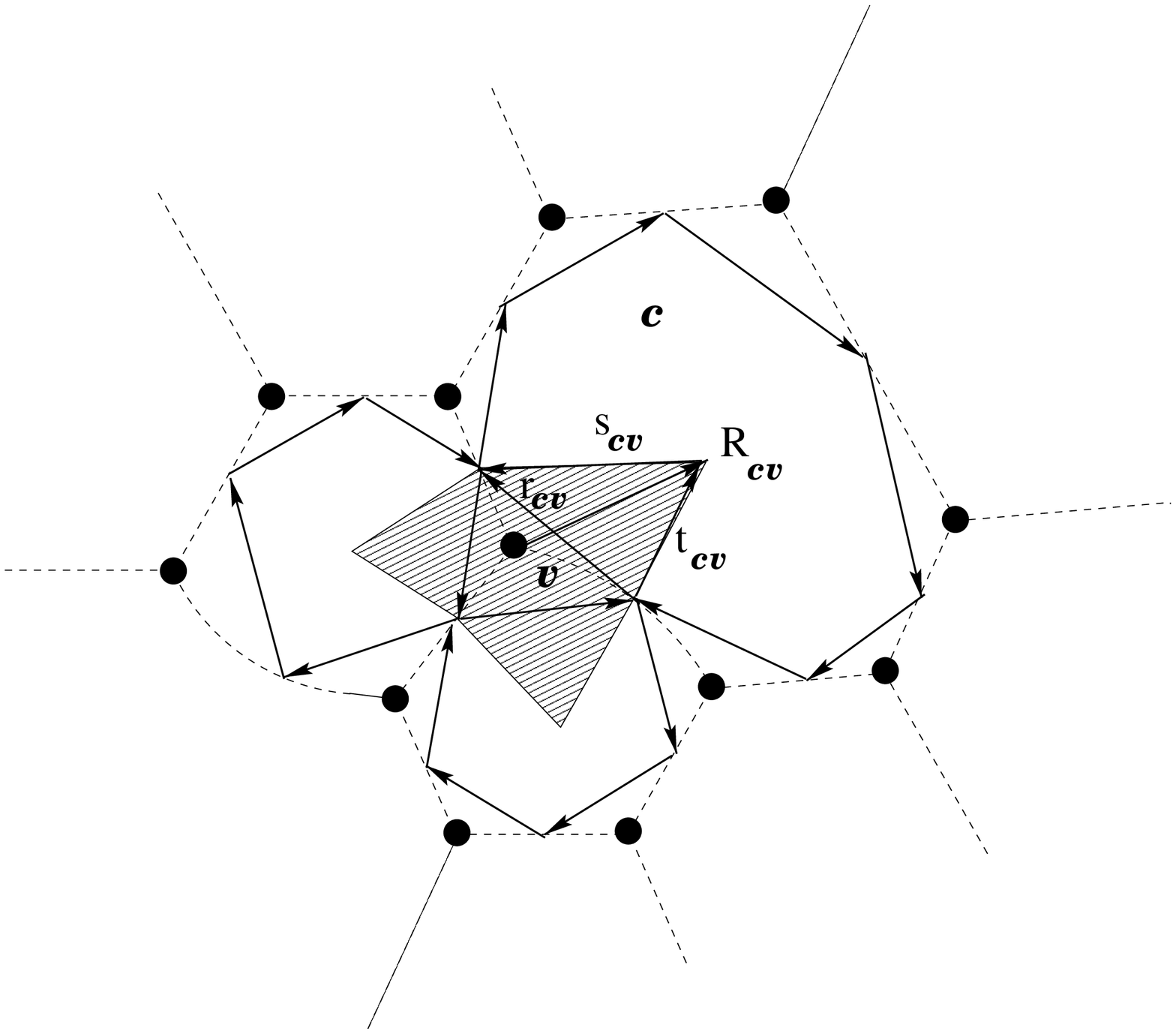,height=6.5cm}}
\caption{
The vectors $\br_{cv}$ and $\brho_{cv}$ shared between vertex $v$ and cell $c$. $\br_{cv}$ connects two neighboring wall midpoints and is one edge in a clockwise-directed triangle around vertex $v$. The vector $\brho_{cv}$ points from the centroid of triangle $v$ to the centroid of cell $c$.}
\end{figure}

\smallskip 
{\bf (a) General networks} 
\smallskip 
 
Starting from the construction shown in figure 1, let us introduce several definitions. First, we assign directions to the sides of the triangles around the vertices so that every triangle circulates its vertex anticlockwise. Each side is now a vector, $\br$, that lies between a particular vertex $v$ and a particular cell $c$ and can be uniquely indexed $\br_{cv}$ (see figure 2). We further define the centroids of the cells as the mean positions of the wall midpoints. Finally, let us define vectors $\bR_{cv}$ that extend from the centroid of triangle $v$ to the centroid of a neighbor cell $c$. The vectors $\br_{cv}$ and $\bR_{cv}$ form two self-dual networks, each edge of which intersects an edge of the other.  
 
Step 1: Each cell is assigned a force $\bff_c$ that acts at the centroid of the cell. It is in terms of the cell forces that we wish to parameterize the contact forces. Every contact sits between two cells, e.g. the contact point between vertices $v$ and $v'$ that is party to cells $l$ and $m$ in figure 3. The force that triangle $v$ exerts on triangle $v'$ is parameterized as  
\begin{equation}
\bff_{vv'} = \bff_l - \bff_m 
\label{eq:Ci}
\end{equation}
The sign convention in eq. (\ref{eq:Ci}) is such that $\bff_c$ is positive when the directed loop formed by the vectors $\br_{cv}$ around cell $c$ is {\it toward} vertex $v'$ at the $vv'$-contact and vice versa for vertex $v$. This parameterization achieves several aims: 
(i) By construction, $\bff_{v'v} = \bff_m - \bff_l = - \bff_{vv'}$ and therefore Newton's third law is satisfied at every contact.
(ii) The forces on each triangle balance automatically. This can be verified by summing all the forces on triangle $v'$, $\sum_v \bff_{vv'}$. Each of the cell forces $\bff_l$, $\bff_m$, and $\bff_n$ contributes to this sum exactly twice, once with a positive and once with a negative sign and therefore this sum vanishes identically. 
(iii) The number of degrees of freedom has been reduced from $6N$ to $2N$, namely, to one force per cell. This coarse-graining corresponds to a reduction in the number of equations since the force balance constraints have been now satisfied. Only one torque balance condition per triangle remains, giving $2N$ equation to determine the unknowns. 
 
\begin{figure} 
\centerline{\psfig{file=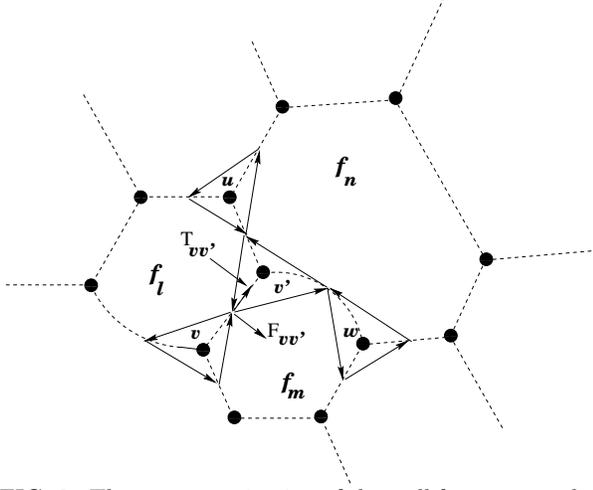,height=6.5cm}} 
\caption{ 
The parameterisation of the wall forces around vertex $v'$ in terms of its 
neighboring cell forces. $\bff_{vv'} = \bff_l - \bff_m$ is the force that 
triangle $v$ exerts on triangle $v'$. The sign convention is that $\bff_l$ is 
positive because the directed loop formed by the vectors $\br$ around cell $l$ 
is {\it toward} triangle $v'$ and vice versa for cell $m$.} 
\end{figure} 

Step 2: We now wish to continue the discrete cell forces into a continuous field, $\bcF$. Regarding every cell force as located at the centroid of its cell, let us interpolate the three forces around every triangle piecewise linearly. The interpolation consists of a planar triangular surface whose corners are located above the centroids of cells $l$, $m$, and $n$, and whose heights are, respectively, $\bff_l$, $\bff_m$, and $\bff_n$. The value of the continuous force field at the vertex $v$, $\bcF_v$, is the weighted mean of the cell forces around it, as illustrated in figure 4 for the $x$-component. In terms of $\bcF$, the cell forces around vertex $v$ can be presented as 

\begin{equation}
\bff_c = \bcF_v + \bR_{cv}\cdot\bnabla\bcF
\label{eq:Ciii}
\end{equation}
  
\begin{figure}
\centerline{\psfig{file=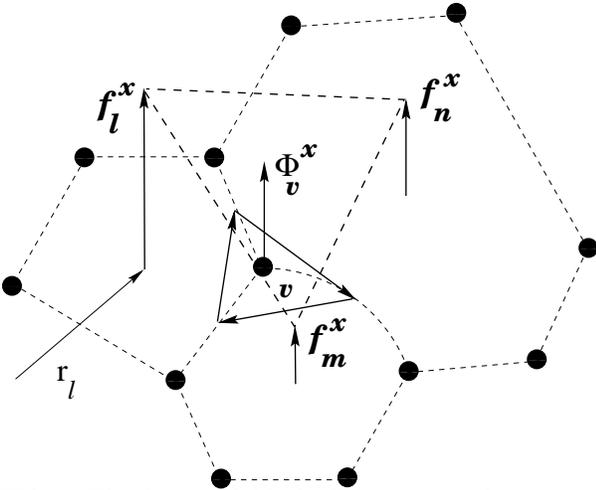,height=6.5cm}}
\caption{
The discrete field of cell forces is made into a continuous force field $\bcF$ by constructing a piecewise linear triangular surface around each vertex. An example of such a surface for the $x$-component of the cell forces around vertex $v$ is shown. The corners of the triangular surface are above the centroids of the cells $l$, $m$ and $n$ that are located at $\br_l$, $\br_m$, and $\br_n$. The heights of the planar surface at these points are, respectively, $\bff_l^x$,  $\bff_m^x$, and $\bff_n^x$. The value of the surface above the vertex gives $\bcF_v$. In terms of the continuous field, the value of any cell force $\bff_c$, immediately neighboring vertex $v$, is exactly $\bff_c = \bcF + \bR_{cv}\cdot\bnabla\bcF$.
}
\end{figure}

Step 3: In terms of the vectors $\br_{cv}$ and forces $\bff_c$, the force moment on triangle $v$ can be written as 
 
\begin{equation}
S_v^{ij} = \sum_c r_{cv}^{\ i} f_c^{\ j}
\label{eq:Civ}
\end{equation}
where the sum runs over all the $c$-cells surrounding grain $v$. Substituting (\ref{eq:Ciii}) into (\ref{eq:Civ}) and observing that $\sum_c \br_{cv} = 0$ gives 
 
\begin{equation}
S_v^{ij} = \sum_{c,k} r_{cv}^{\ i} R_{cv}^k \partial_k \Phi^j  = \left( \hC_v\cdot\bnabla \bcF \right)^{ij}
\label{eq:Cv}
\end{equation}
where

\begin{equation}
\hC_v = \sum_c \br_{cv}\bR_{cv}
\label{eq:Cvi}
\end{equation}
and the component notation has been suppressed. $\hC_v$ is a fabric tensor that characterizes the local geometry around vertex $v$ and plays a crucial role in the following analysis. It is convenient to separate $\hC_v$ into its antisymmetric part $A_v\heps = (\hC_v-\hC_v^T)/2$ and symmetric part $\hP_v = (\hC_v+\hC_v^T)/2$, where $A_v$ is a prefactor and $\matrix{\heps}={0 \,1 \choose -1 \,0 }$ is the $\pi/2$ rotation tensor. Both these parts have direct geometrical interpretations: $A_v=|\br_{cv}\times\bR_{cv}|/2$ is the area associated with vertex $v$ that is shown shaded in figure 2. The entire system can be conveniently tesselated and covered by these areas, namely, $A_{sys} = \sum_v A_v$. This feature is useful for calculations of effective volumes \cite{BE}. To identify the geometrical interpretation of the symmetric part, let us rewrite the vectors $\br_{cv}$ and $\bR_{cv}$ in expression (\ref{eq:Cvi}) in terms of the vectors $\bs_{cv}$ and $\bt_{cv}$ shown in figure 2. It is straightforward to show that $\hP_v$ can be  represented in the form 
 
\begin{equation}
\hP_v = {1\over 2} \sum_c \left( \bs_{cv}\bs_{cv} - \bt_{cv}\bt_{cv}\right)
\label{eq:Cvii}
\end{equation}
The departure of $\hP_v$ from zero around vertex $v$ is a direct measure of how much the $v$ triangle is rotated relative to its immediate environment. For example, if for a particular self-dual pair the vectors $\br$ and $\bR$ are perpendicular then the symmetric part is exactly zero. Thus, this tensor characterizes the {\it rotational} disorder around vertex $v$. Expression (\ref{eq:Cvii}) makes it straightforward to see that summing $\hP_v$ over any patch of the system the contribution from cells fully enclosed inside the patch cancel out and the only non-vanishing contribution comes from boundary terms. Thus, the average of the symmetric part over the entire system vanishes at least as fast as $1/\sqrt{N}$. 
 
Consider now a small patch $\Gamma$ with boundary $\partial\Gamma$ that consists of several cells. The mean stress inside this region is

\begin{equation}
\sig_{\Gamma} = \langle \hS_v \rangle_{\Gamma} / A_{\Gamma} = \langle \hC_v\cdot\bnabla\bcF \rangle_{\Gamma} / A_{\Gamma}
\label{eq:Cviiaa}
\end{equation}
where $\langle \dots \rangle_{\Gamma}$ stands for an area average over the patch $\Gamma$ and $A_{\Gamma}$ is its area. It has been shown in \cite{BB}\ that

\begin{equation}
\langle \hC_v\cdot\bnabla\bcF \rangle_{\Gamma} =  \langle \hC_v \rangle_{\Gamma}\cdot\langle \bnabla\bcF \rangle_{\partial\Gamma}
\label{eq:Cviia}
\end{equation}
where $\langle \dots \rangle_{\partial\Gamma}$ stands for an average over the boundary of $\Gamma$. Combining (\ref{eq:Cviiaa}) and (\ref{eq:Cviia}) we have  
 
\begin{equation}
\langle \bnabla\bcF \rangle_{\partial\Gamma} = \langle \hC_v \rangle_{\Gamma}^{-1} A_{\Gamma} \hsig_\Gamma = \heps^{-1} \hsig_\Gamma 
\label{eq:Cviib}
\end{equation}
where the last equality uses the fact that the average of $\langle \hC_v \rangle_{\Gamma} = A_\Gamma [\heps + \langle P\rangle_{\Gamma}]$ and that the average $\langle P\rangle$ vanishes faster than $1/\sqrt{N}$. We can now apply the yet unused condition of torque balance to a small region in the system: 
 
\begin{equation}
\cA\left\{ \hC_v\cdot\bnabla\bcF \right\} =  \cA\left\{ \hC_v \heps^{-1} \hsig \right\} =  0
\label{eq:Cviic}
\end{equation}
where $\cA\{ \dots \}$ stands for the antisymmetric part of $\{ \dots \}$. Separating $\hC_v$ into its symmetric and antisymmetric parts we obtain two equations: The first,  
 
$$\cA\left\{ \hsig \right\} =  0 $$ 
reassuringly yields the expected condition that the mean stress field is symmetric. The second, 
 
\begin{equation}
\cA\left\{ \hP \heps^{-1} \hsig \right\} = p^{xx}\sig_{yy} + p^{yy}\sig_{xx} -  2p^{xy}\sig_{xy} = 0 
\label{eq:Cviid}
\end{equation}
is new and gives the missing constitutive equation. By defining a rotated version of $\hP$, $\hQ = \heps\hP\heps^{-1}$, this equation can be rewritten more compactly 
 
\begin{equation}
\hQ : \hsig =  0 
\label{eq:Cviii}
\end{equation}
This new constitutive equation couples the stress to local fluctuations in the geometry of the cellular structure and completes the set of equations for the stress field in planar systems. Just like the tensor $\hP$, $\hQ$ characterizes the local rotational disorder and it is this particular aspect of the local geometry that the stress field couples to.  
 
\begin{figure}
\centerline{\psfig{file=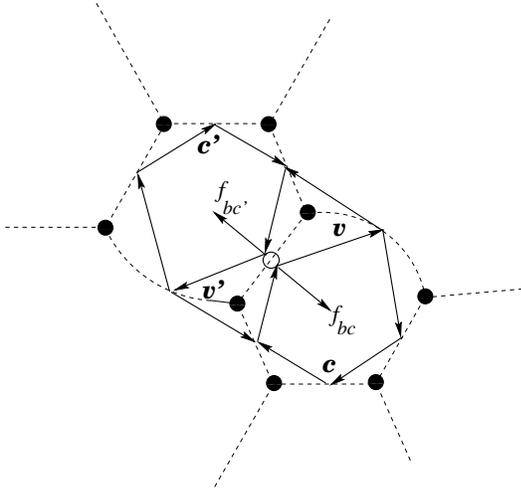,height=6.5cm}}
\caption{
Making all the grains rough and inserting a vanishingly small ball bearing (white circle) 
between touching grains at the contact maintains the original intergranular forces.}
\end{figure}

\smallskip 
{\bf Application to Plateau networks} 
\smallskip 

As formulated above, the theory applies only to general networks whose cell walls meet at arbitrary angles. This is because in general networks there are $6N$ degrees of freedom to determine on the cellular level. This necessitates $2N$ rigid bodies for obtaining equations for the balance of forces and torques, which are conveniently provided by the triangles around the vertices. In contrast, loading a Plateau network gives rise only to $3N$ degrees of freedom on the cellular level: the mean normal force on every cell wall. This then implies that we have to use a mapping to assemblies of smooth grains that are located within the cells and which transmit at each contact only force that is normal to the contact surface. 
The adaptation to Plateau networks goes through the observation \cite{BB} that it is possible to extend the theory to assemblies of smooth grains. This is done via a further mapping from assmeblies of $N$ smooth rigid grains into assemblies of $4N$ rough rigid grains as follows: First, the original smooth grains are assigned an infinite friction coefficient, making them rough and thus, in principle, able to support tangential contact forces. Second, an infinitesimally small circular grain is inserted at each contact point, as shown in figure 5. The function of these additional grains is to act as ball bearings and so transmit only the normal forces between the original grains. This preserves the directions and magnitudes of the forces in the original assmebly (see figure 5). The mean coordination number in the original system is six per grain giving $3N$ contacts altogether. This means that there are $3N$ extra grains and the new assembly consists of $4N$ grains all in all. The new system reproduces the original field of intercellular forces in the smooth system in the limit when the radii of the ball bearings tend to zero. Let us make sure that the number of equations still matches the number of unknowns: In the original system there were $3N$ contacts, each of which corresponding to one unknown. In the 'decorated' system each contact point split into two and so we have $6N$ contacts. Since each ball bearing has two contacts then the force balance equations on them reduce to the condition that the force at one contact point is the same magnitude as the force at the opposite contact, but opposite in direction. This immediately satisfies torque balance on the ball bearing. There remain therefore the $3N$ equations of force and torque balance on the original grains, which exactly take care of the $3N$ unknowns.  
 
Now we have an assembly of rough grains for which the theory is valid and to which equation (\ref{eq:Cviid}) applies. Choosing the radii of all ball bearings to be the same, BB have shown that the equation remains non-trivial when this radius tends to zero. This extends the theory to assemblies of smooth grains and therefore to Plateau networks. 
 
\bigskip 
\ni {\bf V. Conclusion, discussion and future directions} 
\bigskip 
 
This paper has addressed the problem of determining the stress field in planar trivalent cellular solids that are in mechanical equilibrium under an external loading. By comparing the number of degrees of freedom to the number of available equations on the cellular level it has been shown that the intercellular forces can be determined from statics alone, namely, in terms of only the external loading and knowledge of the microstructure of the system. It has been argued then that since the continuous stress field is only a coarse-grained description of the discrete field of intercellular forces and since those can be determined independently of material compliance, then the stress field must be independent of any stress-strain relations. This led to the question what replaces the conventional constitutive equation of elasticity theory in the determination of the stress field. To find the missing equation, a mapping has been constructed between planar trivalent cellular networks and granular assemblies of infinitely rigid grains at the so-called marginal rigidity state. General structures, where cell walls meet at arbitrary angles, map onto assemblies of rough grains that are centered on the vertices of the original cellular network, while Plateau structures, where cell walls meet at angles of $2\pi/3$ everywhere, map onto assemblies of perfectly smooth grains that are enclosed inside the original cells. The mapping enables the application of the Ball-Blumenfeld theory for stress transmission in assemblies of rigid grains to solid foams. In particular, a constitutive equation has been derived using this approach, which couples the stress to the local rotational disorder of the cellular microstructure. This relation replaces the conventional stress-strain relations and redefines the problem of structure-property relationship in these systems.   
The analysis presented in this paper has several fundamental and practical implications, some of which are listed below. 
 
\smallskip 
\noindent {\bf Fundamental issues}:  
\smallskip 
 
\noindent (i) Although the discussion here focused on comparison with elasticity, it should be remembered that {\it all} conventional theories of stress transmission are based on constitutive relations that link the stress to deformation. Elasticity employs strain while elastoplastic and viscoelastic theories are based on relations to strain rate. By resorting to statics alone, the theory developed here is therefore outside the existing paradigm altogether. It has been shown by BB that the new formalism recovers the Airy stress function as the solution for the symmetric divergence-free stress field. The fact that this solution, which is usually obtained within conventional elasticity theory, can be derived without the use of any stress-strain relation clearly demonstrates that these are redundant for determining the stresses that develop in trivalent solid foams. This also means that there is a connection between the two approaches. To see why such a relation must exist suppose we apply an infinitesimally small strain to a solid foam, of the type discussed here, at the end of which the medium settles into the aforementioned isostatic state. Elasticity theory should be able to calculate the stress field just before this state is reached, while the new theory should be able to do so after all deformation has ceased. The stress field cannot change discontinuously at the end of the infinitesimal straining process and therefore both theories must yield the same answer in the limit when the strain goes to zero. This issue is currently under investigation by this author and preliminary results suggest that indeed there is a relation between the two theories.  
 
\noindent (ii) As mentioned already, in the context of granular assemblies, when all the intergranular forces can be determined from statics alone the system is in a marginal rigidity state \cite{BEB} \cite{puna}, which is interpreted as a new state of solid matter. In granular assemblies, the basic condition for being in this state is that the average coordination number per grain is three for rough grains and six for smooth. The state of matter of any system can be determined by its response to external mechanical loading: Gases need to be held together by an external pressure, liquids flow under stress and solids first deform and then transmit the stress statically. The difference between the equations of stress transmission in marginally rigid assemblies and in conventional solids implies different responses to external loading and hence, by definition, that these are in different states of solid matter. In granular systems this new state lies between the fluid and the traditional solid states.  
Adding more contacts between grains necessitates additional constitutive information beyond the knowledge of the structure alone, while a reduction in the number of contacts takes the assembly outside mechanical equilibrium and renders it fluid. This feature of existence between two conventional states appears to have no immediate analog in solid foams. It is possible that such an analog can be found in the context of the rheology of foams close to the so-called jamming transition point. It is interesting to note that in granular systems there is a direct relation between the constitutive equation (\ref{eq:Cviii}) and the dynamics of yield \cite{BBa}. It would be interesting to try to extend this connection to the flow of foams. 
 
\noindent (iii) The basic principle of force determination on the cellular level without resorting to the compliance of the cell walls does not appear to rely heavily on the two-dimensional nature of the system. In fact, it is possible to extend this argument to open-cell foams where each vertex is the meeting point of four edges and six walls, a rather ubiquitous class of structures in nature. This suggests that the stress field in such three-dimensional systems should also be independent of stress-strain relations and therefore that there is a larger class of solid foams that are at a state of marginal rigidity. 
 
\smallskip 
\noindent {\bf Practical issues}: 
\smallskip 
 
\noindent (i) First and foremost, this theory will clearly make it possible to model disordered planar cellular materials to predict stress transmission, extending the current capabilities. In particular, the theory should make it straightforward to predict 'hot spots' of stress concentration in such systems and make useful connections between the occurrence of hot spots and the  microstructure. This knowledge is expected to both improve prediction of failure in advanced materials and set new goals for the design of microstructural characteristics that prevent such vulnerability.   
 
\noindent (ii) The formal mapping between granular assemblies and cellular systems can be used to transfer results between the two fields. Granular assemblies can only approach the marginal rigidity state under careful consolidation processes \cite{BEB}. In contrast, it is quite common for the dynamics of formation of foams to give rise to triple junctions and so drive the structure naturally to this state. Morever, once a foam has solidified its structure is set, while granular assemblies are fragile by the very nature of this state. This difference means that most measurements in granular assemblies are plagued by large fluctuations, while foams are amenable to more conclusive experimental observations. Therefore, this mapping creates a formal gateway between the two systems and makes it possible to carry out experiment on one type of systems in order to learn about the other.    
 
\bigskip
{\bf Acknowledgements:}
\smallskip

\ni It is a pleasure to acknowledge discussions with Prof R. C. Ball, Prof Sir S. F. Edwards and Dr D. V. Grinev. I am grateful to Prof M. E. Cates who stressed to me the importance of assuming the limit of infinitely thin cell walls to this analysis.


\begin{thebibliography}{99}
\bibitem{sixedge} This is a direct reult of applying Euler's theorem relating the number of
cells, faces and edges to a planar structure. The theorem has been proven by Euler circa 1750.
\bibitem{deform} In fact, this statement extends to all theories that use deformation-stress relationship as constitutive input to solve for the stress field
\bibitem{linetopoint} Although the normal forces are distributed all along the cell walls, it is possible to show that these can be uniquely represented by a localized force $F_{vv'}$ normal to the cell walls. Placing these forces at the midpoints of the walls is assumed here to introduce only a negligible error
\end{thebibliography}
\end{document}